\documentclass[10pt,leqno]{amsart}

\usepackage[T1]{fontenc}
\usepackage{graphicx}
\usepackage{amssymb,amsthm,amsmath}
\usepackage{xcolor,hyperref}
\usepackage{indentfirst,csquotes}

\usepackage{enumerate}
\usepackage{listings}
\usepackage{subcaption}
\usepackage{float}

\hypersetup{
  colorlinks=true,
  linkcolor=black,
  filecolor=black,
  urlcolor=black,
  citecolor=black
}

\lstdefinelanguage{Turtle}{
    alsoletter={:},
    keywords={@prefix,a},
    sensitive=false,
    morecomment=[l]{\#},
    morestring=[b]"
}

\lstdefinelanguage{Jinja}{
    morecomment=[l]{\#},
    morestring=[b]",
    alsoletter={\{\}},
    sensitive=true
}

\title[A Pythonic Functional Approach for Semantic Data Harmonisation]{A Pythonic Functional Approach for Semantic Data Harmonisation in the ILIAD Project}

\author[E.\,Nystad and F.\,Mart\'in-Recuerda]{Erik Johan Nystad and Francisco Mart\'in-Recuerda}
\thanks{SINTEF Digital, Oslo, Norway. \texttt{\{erik.nystad, francisco.martin-recuerda\}@sintef.no}}

\begin{document}
\maketitle

\begin{abstract}
Semantic data harmonisation is a central requirement in the ILIAD project, where heterogeneous environmental data must be harmonised according to the Ocean Information Model (OIM), a modular family of ontologies for enabling the implementation of interoperable Digital Twins of the Ocean. Existing approaches to Semantic Data Harmonisation, such as RML and OTTR, offer valuable abstractions but require extensive knowledge of the technical intricacies of the OIM and the Semantic Web standards, including namespaces, IRIs, OWL constructors, and ontology design patterns. Furthermore, RML and OTTR oblige practitioners to learn specialised syntaxes and dedicated tooling. Data scientists in ILIAD have found these approaches overly cumbersome and have therefore expressed the need for a solution that abstracts away these technical details while remaining seamlessly integrated into their Python-based environments. To address these requirements, we have developed a Pythonic functional approach to semantic data harmonisation that enables users to produce correct RDF through simple function calls. The functions, structured as Python libraries, encode the design patterns of the OIM and are organised across multiple levels of abstraction. Low-level functions directly expose OWL and RDF syntax, mid-level functions encapsulate ontology design patterns, and high-level domain-specific functions orchestrate data harmonisation tasks by invoking mid-level functions. According to feedback from ILIAD data scientists, this approach satisfies their requirements and substantially enhances their ability to participate in harmonisation activities. In this paper, we present the details of our Pythonic functional approach to semantic data harmonisation. We demonstrate its applicability within the ILIAD Aquaculture pilot, where heterogeneous environmental datasets must be integrated and analysed to detect salmon-lice exposure and support farm-level decision-making.

\end{abstract}

\bigskip

\section{Introduction}
\label{Introduction}

The ILIAD project \cite{ILIAD2025} aims to implement an infrastructure for developing interoperable Digital Twins that support different environmental aspects of the ocean.
To validate ILIAD objectives, 12 diverse pilots were defined, including pilots to study oil spills, jelly fish swarms, and environmental conditions of fish farms. A central requirement for enabling interoperable Digital Twins is to implement an effective approach for harmonising heterogeneous environmental data, originating from sensors, numerical models, registers, and public APIs, into a common, semantically rich representation. To accomplish this objective, ILIAD has adopted W3C Semantic Web open standards \cite{w3c_sw} and implemented the Ocean Information Model (OIM) \cite{OIM2025}, a modular family of ontologies designed to provide a shared conceptual framework for ocean-related information and to ensure semantic interoperability across all ILIAD Digital Twins. OIM builds on established standards and ontologies, such as SOSA \cite{janowicz2019sosa}, QUDT \cite{qudt}, and DCAT \cite{dcat2019}. ILIAD also implemented domain-specific ontologies for biodiversity, geospatial features, aquaculture, and oceanography.

Although OIM provides the required semantic foundation, applying it in practice presents significant challenges for data scientists working in ILIAD pilots. Much of the data arrives in JSON, CSV, or other tabular formats, and analysts must translate simple, domain-level statements, such as “the sea temperature at this farm was 9.4 °C” or “the current velocity at this location was 0.54 m/s”, into structured RDF involving SOSA observations, QUDT units, OIM observable properties, features of interest, and persistent URIs.

Several approaches have been developed to support semantic data harmonisation, including declarative mapping languages such as RML \cite{dimou2014rml} and templating languages such as OTTR \cite{skjaeveland2024reasonable}. While these frameworks provide valuable abstractions for simplifying harmonisation tasks, they nonetheless require substantial familiarity with Semantic Web standards, including namespaces, IRIs, OWL constructors, and ontology design patterns. In addition, both RML and OTTR demand that practitioners learn specialised syntaxes and dedicated tooling. Within ILIAD, data scientists have found these requirements overly cumbersome and have therefore expressed the need for solutions that hide such technical complexity while integrating naturally into their Python-based workflows. Although Python libraries exist for instantiating OTTR templates, such as maplib \cite{bakken2023maplib} and pyOTTR \cite{pyottr}, they remain grounded in the OTTR specification and thus inherit the same fundamental challenges.

In response to the feedback provided by ILIAD data scientists, we have developed a Pythonic functional approach to semantic data harmonisation that enables users to produce correct RDF through simple function calls. The functions, structured as Python libraries, encode the design patterns of the OIM and are organised across multiple levels of abstraction.  Low-level functions directly expose OWL and RDF syntax, mid-level functions encapsulate ontology design patterns, and high-level functions orchestrate data harmonisation tasks by invoking mid-level functions. Some ontologies such as QUDT, might require the implementation and maintenance of thousands of low-level functions referring to specific units of measure or quantity kinds. To automate the generation of these functions, we implement a data-driven approach based on Jinja \cite{jinja} templates instantiated using the results of SPARQL queries to authoritative vocabularies. Notice that we envision that the implementation of low-level and mid-level functions, are done by ontology engineers, whereas high-level functions are created and maintained by data scientists. This is a task data scientists found easier to accomplish, enhancing their ability to participate in data harmonisation activities.

In this paper, we present the details of our Pythonic functional approach to semantic data harmonisation. We demonstrate its applicability within the ILIAD Aquaculture pilot, where heterogeneous environmental datasets must be integrated and analysed to detect salmon-lice exposure and support farm-level decision-making. A simplified version of the implementation we created in the ILIAD project is available in a dedicated open GitHub repository\footnote{\url{https://github.com/SINTEF/iliad-pythonic-harmonisation-demo}}.

The paper is organised as follows. Section \ref{use_case} introduces the Aquaculture pilot and discusses our attempts to use common approaches for semantic data harmonisation, including RML and OTTR. Section \ref{pythonic_approach} describes our Pythonic approach for semantic data harmonisation. Section \ref{related_work} discusses relevant related work. Finally, the main conclusions of the paper and future work are presented in Section \ref{conclusions_and_future_work}. 

\section{Harmonising Data in the ILIAD Aquaculture Pilot}
\label{use_case}
\begin{figure}[t]
\centering
\includegraphics[width=0.4\textwidth]{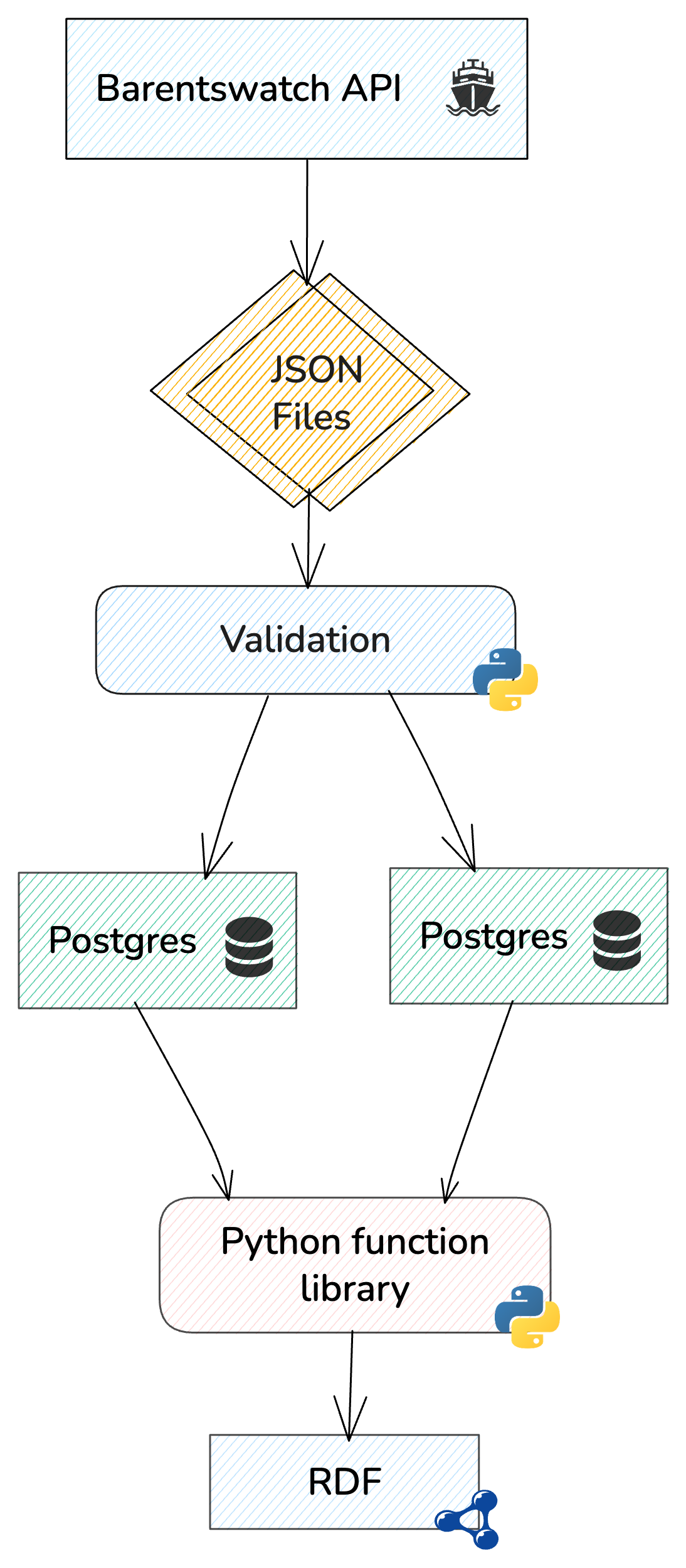}
\caption{Data-ingestion and harmonisation pipeline in the ILIAD Aquaculture pilot.}
\label{fig:barentswatch_pipeline}
\end{figure}

\begin{figure}[h!]
\centering
\includegraphics[width=\textwidth]{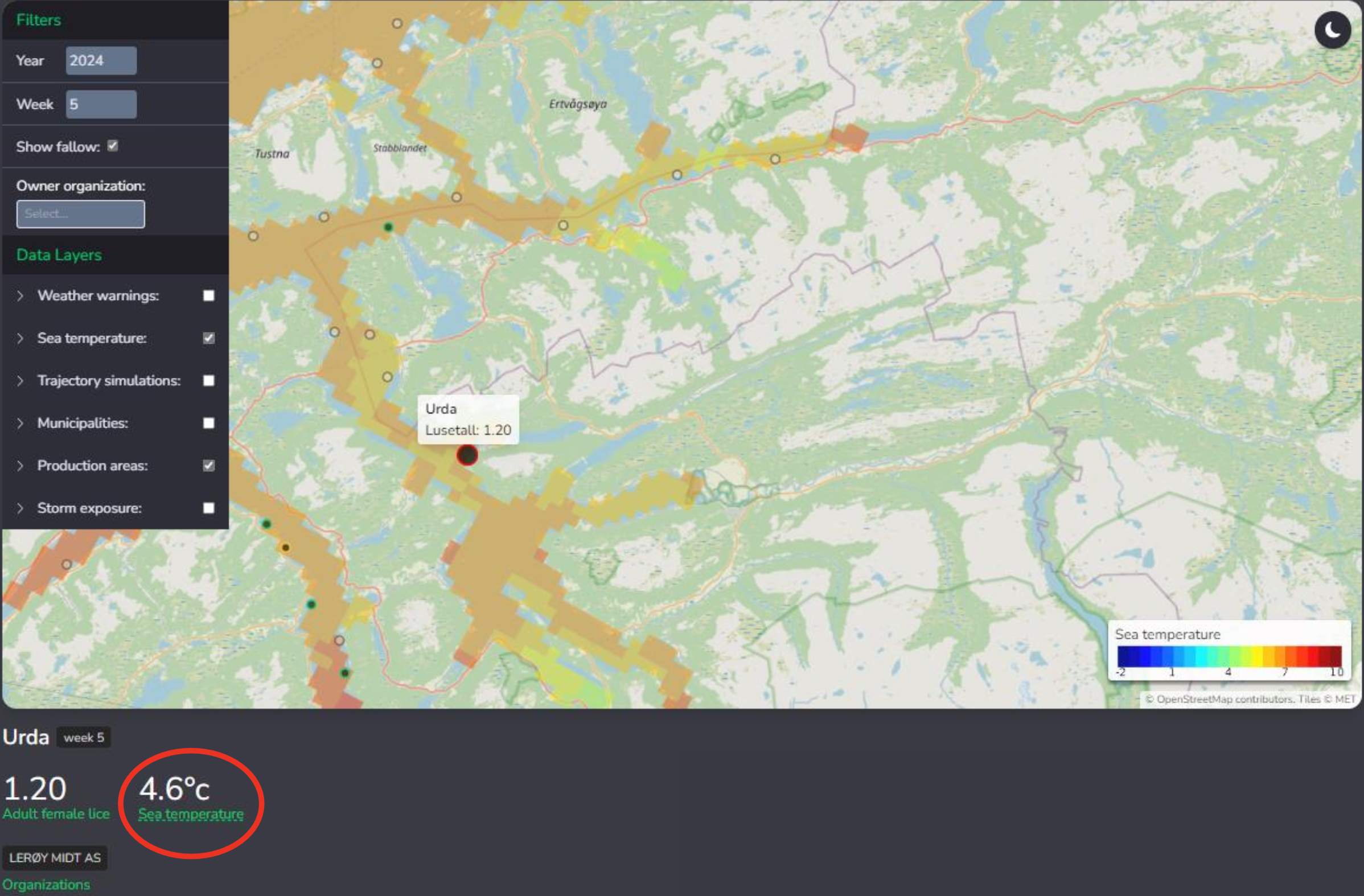}
\caption{ILIAD Dashboard displaying a sea temperature measurement at a specific time and location.}
\label{fig:dashboard}
\end{figure}

Aquaculture operators in Norway are required by law to report salmon lice counts, a regular parasite that affects fish welfare and fish health, observed water temperature and infections to the food authority once per week. The Aquaculture pilot requests information on lice counts and environmental conditions from the service BarentsWatch \cite{barentswatch}. 
The pilot runs 24-hour trajectories with OpenDrift \cite{dagestad2018opendrift} from all locations based on the forecast of the operational ocean model Norkyst800. Figure \ref{fig:barentswatch_pipeline} provides a simplified view of the data ingestion and harmonisation pipeline in the ILIAD Aquaculture pilot. Data is periodically collected from BarentsWatch API and validated before being stored in Postgres. The data is transformed into OIM-compliant RDF graphs that are submitted to the ILIAD data platform. Figure \ref{fig:dashboard} illustrates how information is visualized to enable decision support in the Aquaculture pilot. The figure specifically shows sea temperature measurements.

Following the ILIAD design requirements, all data collected and produced by the Aquaculture pilot must be harmonised using the Ocean Information Model (OIM). Early in the project, we evaluated several established tools and methodologies, including RML and OTTR. This section summarises that evaluation process. Rather than providing an abstract comparison of tools, we illustrate the discussion through the concrete transformation of the following minimal sea-temperature record into semantically enriched, OIM-compliant RDF:
\begin{lstlisting}[language=Python,basicstyle=\ttfamily\small]
id = 1234
sea_temperature_celsius = 4.6
timestamp   = "2025-06-27T01:00:00Z"
latitude    = 70.41
longitude   = 0.00
\end{lstlisting}

Even this single value requires a standardised modelling pattern involving a SOSA Observation, which includes a feature of interest representing the geographic location, and an observable property related to a sea temperature measurement. Following best practices from OIM, the quantity value, the unit of measure, and the quantity kind are represented using QUDT. The previous example can be specified in RDF as follows:

\begin{lstlisting}[language=Turtle,basicstyle=\ttfamily\small]
oim-obs:sea_temperature_1234 a sosa:Observation;
  sosa:hasFeatureOfInterest oim-feat:loc_70.41_0.0;
  sosa:hasResult oim-res:sea_temperature_1234;
  sosa:observedProperty oim-prop:seaTemperature;
  sosa:resultTime "2025-06-27T01:00:00Z"^^xsd:dateTime .

oim-res:sea_temperature_1234 a qudt:QuantityValue;
  qudt:numericValue "4.6"^^xsd:float;
  qudt:unit unit:DEG_C .
\end{lstlisting}

To produce the RDF shown earlier, we experimented with the declarative mapping language RML, which requires the definition of multiple Triples Maps. A Triples Map specifies how to generate RDF triples that share a common subject \cite{dimou2014rml}. In our case, a separate Triples Map was needed for each SOSA observation and each SOSA result. Creating these maps involves selecting the appropriate IRIs, classes, and properties defined in OIM. We evaluated several tools that support RML-based harmonisation of environmental data, including RMLMapper \cite{Heyvaert_RMLMapper-JAVA} and YARRRML Parser \cite{heyvaert2018declarative}, the latter providing a more concise YAML-based syntax. Figure \ref{fig:rml_example} presents a possible RML implementation for SOSA observations and results, with some names and IRIs abbreviated for clarity.

We introduced RML syntax and its supporting tools to the data scientists in ILIAD, but they found the approach challenging to learn. The difficulty extended beyond understanding the syntax and tooling: effective use of RML also required substantial familiarity with the OIM, including its ontologies, design patterns, namespaces, IRIs, and OWL constructors. Additionally, they expressed concerns about the implementation, reuse, and long-term maintenance of large numbers of Triples Maps. RML offers limited mechanisms for reuse, as each new observable property must be mapped almost entirely from scratch, even when the underlying SOSA design pattern is identical. These limitations prompted us to explore more modular solutions, leading us to investigate OTTR, the Reasonable Ontology Templates framework.

\begin{figure*}[t]
\centering
\begin{minipage}[t]{0.49\textwidth}
\begin{lstlisting}[language=Turtle,basicstyle=\ttfamily\footnotesize]
:SeaTempObs
  a rr:TriplesMap ;
  rml:logicalSource :SeaTempSource ;

  rr:subjectMap [
    rr:template ".../obs/sea_temp_{id}" ;
    rr:class sosa:Observation
  ] ;

  # observed property
  rr:predicateObjectMap [
    rr:predicate sosa:observedProperty ;
    rr:object   oim-prop:seaTemperature
  ] ;

  # feature of interest
  rr:predicateObjectMap [
    rr:predicate sosa:hasFeatureOfInterest ;
    rr:objectMap [
      rr:template ".../feat/loc_{lat}_{lon}"
    ]
  ] ;

  # link to result node (join by id)
  rr:predicateObjectMap [
    rr:predicate sosa:hasResult ;
    rr:objectMap [
      rr:parentTriplesMap :SeaTempRes ;
      rr:joinCondition [ 
        rr:child "id" ; 
        rr:parent "id" 
      ]
    ]
  ] ;

  # result time from timestamp
  rr:predicateObjectMap [
    rr:predicate sosa:resultTime ;
    rr:objectMap [
      rml:reference "timestamp" ;
      rr:datatype xsd:dateTime
    ]
  ] .


\end{lstlisting}
\end{minipage}\hfill
\begin{minipage}[t]{0.49\textwidth}
\begin{lstlisting}[language=Turtle,basicstyle=\ttfamily\footnotesize]

:SeaTempRes
  a rr:TriplesMap ;
  rml:logicalSource :SeaTempSource ;

  rr:subjectMap [
    rr:template "results/sea_temp_{id}" ;
    rr:class qudt:QuantityValue
  ] ;

  # numeric value: sea_temperature_celsius
  rr:predicateObjectMap [
    rr:predicate qudt:numericValue ;
    rr:objectMap [
      rml:reference "sea_temp_celsius" ;
      rr:datatype xsd:float
    ]
  ] ;

  # unit: degrees Celsius
  rr:predicateObjectMap [
    rr:predicate qudt:unit ;
    rr:object   unit:DEG_C
  ] .


:SeaTempSource a rml:LogicalSource ;
    rml:source             "sea_temp.json" ;
    rml:referenceFormulation ql:JSONPath ;
    rml:iterator           "$[*]" . 


\end{lstlisting}
\end{minipage}

\caption{Example of RML mappings for sea-temperature observations and results.}
\label{fig:rml_example}
\end{figure*}

OTTR is a framework for defining and instantiating OWL ontology design patterns, supported by a family of template languages and associated tools. Unlike RML, which specifies mappings between JSON fields and RDF predicates, OTTR takes a template-based approach. Figure \ref{fig:ottr-bottr-overall} illustrates how OTTR can be applied to our sea-temperature example. We begin by defining a template for a SOSA observation, followed by an InstanceMap that assigns values from the source data to the template’s parameters. A simplified version of both the template and the corresponding InstanceMap is shown in Figures \ref{fig:ottr-template} and \ref{fig:bottr-map}, respectively.

\begin{figure}[t]
\centering

\begin{subfigure}[t]{0.49\linewidth}
\centering
\begin{minipage}[b]{\linewidth}
\begin{lstlisting}[language=Turtle,basicstyle=\ttfamily\footnotesize]
:SeaTempObservation[
    ottr:IRI     ?obs,
    ottr:IRI     ?foi,
    ottr:IRI     ?res,
    xsd:float    ?temp,
    xsd:dateTime ?time
] :: {
    # SOSA observation
    ottr:Triple(?obs, rdf:type,
        sosa:Observation),
    ottr:Triple(?obs,
        sosa:observedProperty,
        oim-prop:seaTemperature),
    ottr:Triple(?obs,
        sosa:hasFeatureOfInterest,
        ?foi),
    ottr:Triple(?obs,
        sosa:hasResult, ?res),
    ottr:Triple(?obs,
        sosa:resultTime, ?time),
    # SOSA result
    ottr:Triple(?res, rdf:type,
        qudt:QuantityValue),
    ottr:Triple(?res,
        qudt:numericValue, ?temp),
    ottr:Triple(?res,
        qudt:unit, unit:DEG_C)
} .
\end{lstlisting}
\end{minipage}
\caption{OTTR template for sea-temperature.}
\label{fig:ottr-template}
\end{subfigure}
\hfill
\begin{subfigure}[t]{0.49\linewidth}
\centering
\begin{minipage}[b]{\linewidth}
\begin{lstlisting}[language=Turtle,basicstyle=\ttfamily\footnotesize,escapechar=\&]
[] a ottr:InstanceMap ;
   ottr:template
       ex:SeaTempObservation ;
   ottr:source
       [ a ottr:H2Source ] ;
   ottr:query """
    SELECT '.../oim/obs/sea_temp_'
      || CAST(id AS VARCHAR) AS obs,
      '.../oim/feature/loc_'
      || CAST(lat AS VARCHAR)
      || '_'
      || CAST(lon AS VARCHAR) AS foi,
      '.../oim/res/sea_temp_'
      || CAST(id AS VARCHAR) AS res,
      sea_temp_celsius, timestamp
    FROM CSVREAD(
      '@@THIS_DIR@@/sea_temp.csv');
   """ ;
   ottr:argumentMaps (
     [ ottr:type ottr:IRI ]      # obs
     [ ottr:type ottr:IRI ]      # foi
     [ ottr:type ottr:IRI ]      # res
     [ ottr:type xsd:float ]     # temp
     [ ottr:type xsd:dateTime ]  # time
   ) .
\end{lstlisting}
\end{minipage}
\caption{InstanceMap for sea-temperature.}
\label{fig:bottr-map}
\end{subfigure}

\caption{Example of OTTR template for sea-temperature observation.}
\label{fig:ottr-bottr-overall}
\end{figure}

Our work with OTTR in the ILIAD project highlighted several aspects that initially made the technology appealing. OTTR’s template system brings a clear modelling discipline that is often missing in ad-hoc RDF generation. By defining reusable templates for recurring semantic structures, such as SOSA observations or QUDT quantity values, OTTR makes the underlying patterns explicit and easier to understand. In addition, OTTR promotes separation of concerns. Ontology engineers can focus on defining robust templates, while general users only need to instantiate them using instance maps. For organisations that already rely on tabular data pipelines or maintain strong ontology-engineering practices, this approach can be highly effective.

However, when we applied OTTR in the ILIAD scientific workflows, several issues emerged. The first challenge was the learning curve. Although OTTR is conceptually elegant, the various syntaxes and tools makes the framework complex for data scientists in ILIAD. These users typically work in Python-based notebooks and import various libraries containing definitions of classes and functions. Even the use of Python libraries for interpreting OTTR templates, such as maplib \cite{bakken2023maplib}, were not well-accepted by ILIAD data scientists. They demanded a pure Python approach where no new syntaxes or tools are required. In response, we designed a more functional Pythonic approach, where the modelling patterns themselves are expressed as ordinary Python functions without an intermediate template language. This approach is presented in the next section.

\section{A Pythonic Template-Based Approach to Semantic Data Harmonisation}
\label{pythonic_approach}

Instead of describing patterns in a separate OTTR file or an RML mapping file, our approach encodes the design pattern of an ontology directly as executable Python. This allows the harmonisation logic to live in the same place as the data-wrangling logic, without the need of context switching, and learning dedicated syntaxes. 

During the design of our Pythonic approach for data harmonisation, we considered two programming paradigms: object-oriented programming and functional programming. The former specifies programs by instantiating objects and orchestrating their interactions \cite{hudak1989conception}. The latter defines programs by applying and composing functions \cite{kindler2011object}. We chose the functional programming paradigm because it has several advantages. For instance, it enhances readability of code, because functions are usually small and contain immutable data structures, where the state of the data does not change. This paradigm also decreases duplicated code by expressing repeated actions using high-level verbs. Based on our experience in the ILIAD project and feedback from participating data scientists, functional programming reduces the learning curve for data scientists familiar with mathematical functions. Finally, it complies with the notion of ontology templates (e.g., OTTR) for ontology design patterns which we found quite appealing. 

\begin{figure}[h!]
\centering
\includegraphics[scale=0.075]{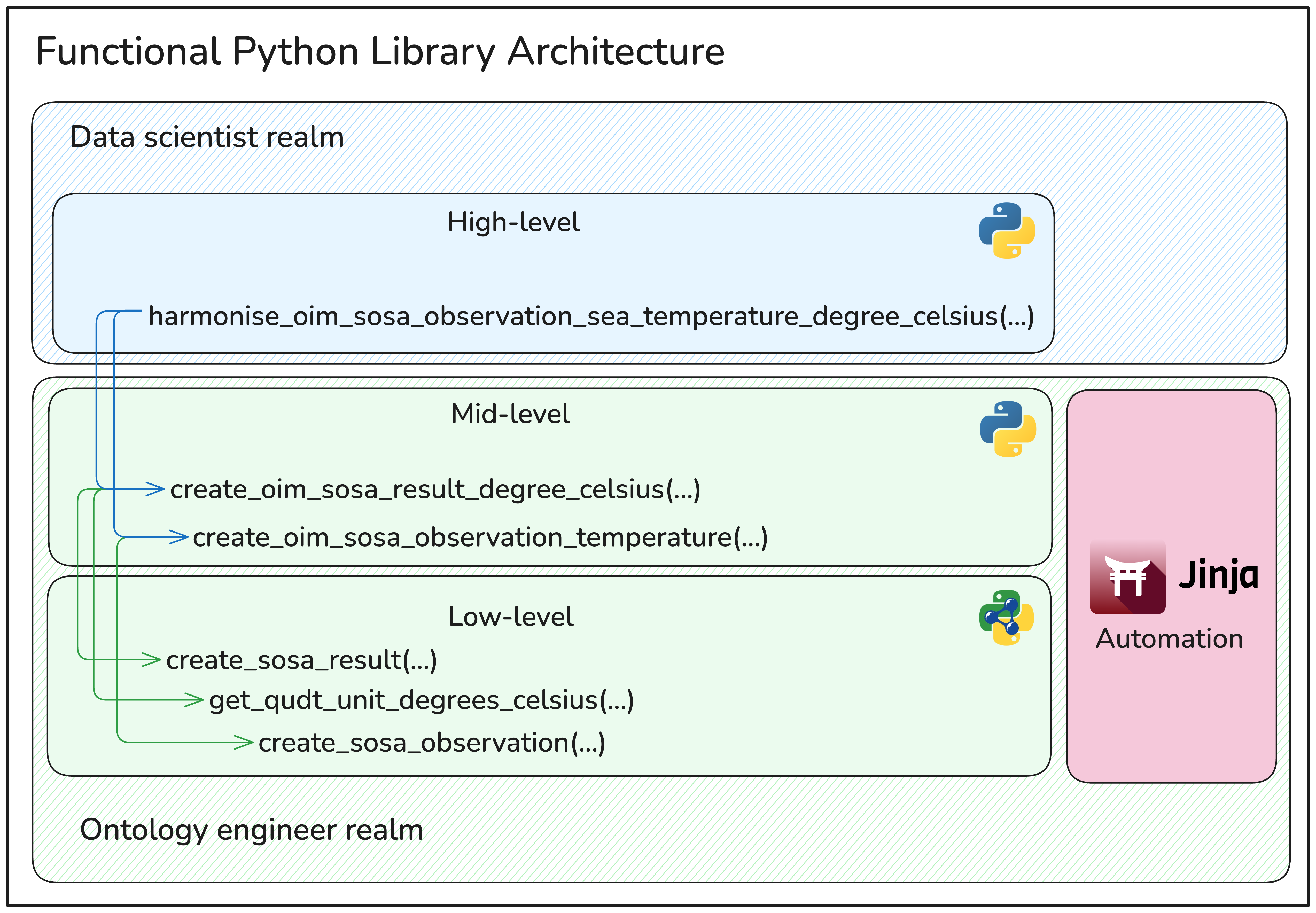}
\caption{Diagram of a hierarchy of Python functions.}
\label{fig:hierarchy_python_functions}
\end{figure}

In the ILIAD project, we tried to distribute the efforts of harmonising data between ontology engineers and data scientists. The formers should be familiar with technical details related to OIM and Semantic Web standards. For the latter, we tried to hide all these technical details. To achieve this objective, we organised functions into different levels of abstraction. Low-level functions directly expose OWL and RDF syntax. Mid-level functions define ontology templates representing ontology design patterns. Finally, high-level functions orchestrate domain-specific data harmonisation tasks by invoking mid-level functions. Ontology engineers are responsible for the implementation and maintenance of Python libraries composed by low-level and mid-level functions. Data scientists are responsible for the implementation and maintenance of high-level functions, and for the specification of data pipelines that collect source data and invoke high-level functions for harmonising this data. Figure \ref{fig:hierarchy_python_functions} depicts the hierarchical structure of the Python functions and clarifies the user roles associated with each level.

As a result of the previous discussion, we defined five main design principles for our Pythonic approach for semantic data harmonisation.

\begin{enumerate}
    \item \textbf{Functional programming for encoding ontology templates.}  
    Ontology template functions are defined according to core functional programming principles, such as immutability, composability, and the use of small, reusable functions.
    
    \item \textbf{Ontology template functions reflect domain concepts.}  
    A sea temperature observation becomes a function with parameters that data 
    scientists already expect: values, timestamps, and feature identifiers.

    \item \textbf{Composability mirrors OTTR’s template nesting.}  
    High-level functions orchestrate domain-specific data-harmonisation tasks by calling mid-level ontology-template functions, which in turn rely on low-level functions that expose OWL and RDF syntax.

    \item \textbf{Different levels of abstraction for different users.}  
    Non-experts on Semantic Web technologies only create, maintain and apply high-level functions that hide technical details encoded in mid-level and low-level functions, implemented by ontology engineers.
    
    \item \textbf{Ontology template functions are packaged as Python libraries.}  
    Functions are organised and distributed as Python libraries, which facilitates testing, debugging, and integration with existing Python workflows.

\end{enumerate}

 In the next section, more details are provided on how ontology design patterns can be represented by composable python functions.

\subsection{Different Levels of Abstraction of Python Functions}
This section provides more details about how we organise python functions on
different levels of abstraction. As discussed earlier, high-level functions are responsible for orchestrating domain-specific data harmonisation tasks by invoking mid-level functions, and they are designed to hide technical details to non-experts on Semantic Web technologies. To illustrate this, consider again our simplified sea temperature example. A call for a high-level function that returns an RDF graph, represented by the variable $g$, could be specified as shown at the top in Figure \ref{fig:high-level-function}. Notice that we use descriptive function and variable names to facilitate understandability for non-expert users and Foundational Language Models.

This functional style mirrors OTTR's approach of template instantiation, but in a Python ecosystem familiar to data scientists. Behind the scenes, this high-level function calls two mid-level functions, one for creating the SOSA observation, and the other for creating the SOSA result. The output of the mid-level functions are RDF graphs that are combined to produce the final output of the high-level function. In practice, the corresponding high-level function is presented at the bottom in Figure \ref{fig:high-level-function}.

\begin{figure}[t]
\centering
\begin{lstlisting}[language=Python,basicstyle=\ttfamily\footnotesize]
# Call to a high-level harmonisation function for sea temperature 
# in degree celsius.
g += harmonise_oim_sosa_observation_sea_temperature_degree_celsius(
    quantity_value=sea_temperature,
    observation_id=observation_id,
    result_time=timestamp,
    feature_of_interest_id=f"loc_{lat}_{lon}",
)


# Implementation of a high-level function for sea temperature in 
# degree celsius.
def harmonise_oim_sosa_observation_sea_temperature_degree_celsius(
    sea_temperature: float,
    observation_id: str,
    result_time: datetime,
    feature_of_interest_id: str,
) -> Graph:

    obs_iri, obs_graph = create_oim_sosa_observation_sea_temperature(
        observation_id=observation_id,
        result_time=result_time,
        feature_of_interest_id=feature_of_interest_id
    )

    result_iri, result_graph = create_oim_sosa_result_temperature_degree_celsius(
        measured_value=sea_temperature,
        observation_iri=obs_iri
    )

    return obs_graph + result_graph
\end{lstlisting}

\caption{Example of a high-level harmonisation function for sea temperature.}
\label{fig:high-level-function}
\end{figure}

Technical details about how SOSA observations and results are defined and serialised as RDF graphs are hidden in the implementation of mid-level and low-level functions. Data scientists are only concerned with combining mid-level functions to produce high-level functions that are executed to harmonise their data. This proved decisive in ILIAD: the moment the semantic templates became Python functions, the barrier between “semantic modelling” and “data processing” dissolved. Instead of learning a new language, users interacted with familiar function signatures, received Python type errors instead of template-engine failures, and integrated harmonisation directly into existing pipelines without converting formats or maintaining separate mapping layers. This functional template approach therefore represents the point in our journey where semantic interoperability became practically achievable within the ILIAD pilots. It preserved the modelling discipline we appreciated in OTTR, but delivered it in the environment the data scientists were already using, namely Python.

In the implementation of the example of high-level functions for harmonising sea temperature measurements, two mid-level functions were called: 
\begin{itemize}
    \item \texttt{create\_oim\_sosa\_observation\_sea\_temperature(...)}
    \item \texttt{create\_oim\_sosa\_result\_temperature\_degree\_celsius(...)}
\end{itemize}

A possible implementation of the latter function is depicted in Figure \ref{fig:mid-level-function}.
\newline

\begin{figure}[h]

\centering
\begin{lstlisting}[language=Python,basicstyle=\ttfamily\footnotesize]
def create_oim_sosa_result_sea_temperature_degree_celsius(
    sea_temperature: float, 
    observation_iri: URIRef
) -> Graph:
    unit = get_qudt_unit_degree_celsius()
    oim_sosa_result_iri = create_oim_sosa_result_iri()

    oim_sosa_result_graph = create_sosa_result(
        sea_temperature, observation_iri, unit, oim_sosa_result_iri
    )
    return oim_sosa_result_iri, oim_sosa_result_graph
\end{lstlisting}

\caption{Example of a mid-level harmonisation function for sea temperature.}
\label{fig:mid-level-function}
\end{figure}

We observe that several technical details become visible. For example, the code explicitly references IRIs defined using the URIRef type provided by the RDFLib library \cite{Krech_RDFLib_2025}. The logic of the mid-level function is implemented by invoking three specific low-level functions. The first retrieves the IRI corresponding to the unit of measure degrees Celsius defined by the ontology QUDT. The second constructs the IRI for a SOSA result\footnote{In this example, a SOSA result has an IRI, but it can also be specified using a blank node.}. The third generates the RDF representation of that SOSA result. In this way, repetitive logic for obtaining or constructing IRIs and RDF graphs is encapsulated within dedicated low-level functions. A possible implementation of the low-level functions discussed earlier is presented in Figure \ref{fig:python-low-level-functions}. Notice that the function \texttt{create\_sosa\_result(...)} utilises the RDFLib Python library for generating RDF graphs. We note that the use of mutable \texttt{Graph} objects within these low-level functions means the implementation is not purely functional. This is a deliberate trade-off: RDFLib provides mature and well-tested RDF serialisation capabilities, and wrapping it in a strictly functional interface would have added complexity without clear practical benefit. The functional discipline is maintained at the mid- and high-level layers, where functions compose immutable outputs, while the low-level layer leverages RDFLib's imperative API for pragmatic reasons. 

\begin{figure*}[t]
\centering

\begin{minipage}[t]{0.5\textwidth}
\begin{subfigure}[t]{\linewidth}
\begin{minipage}{\linewidth}
\begin{lstlisting}[language=Python,basicstyle=\ttfamily\scriptsize]
def create_sosa_result(
    measured_value: float,
    observation_uri: URIRef,
    unit: URIRef,
    result_uri: URIRef,
) -> Graph:
    g = Graph()

    # Add the result triples to the graph
    g.add((result_uri, RDF.type, SOSA.Result))
    g.add((result_uri, SOSA.hasValue,
           Literal(measured_value, datatype=XSD.float)))
    g.add((result_uri, SOSA.hasUnit, unit))

    # Add as a qudt quantity value
    g.add((result_uri, RDF.type, QUDT.QuantityValue))

    # Connect the result to the observation
    g.add((observation_uri, SOSA.hasResult,
           result_uri))

    return g
\end{lstlisting}
\end{minipage}
\caption{Generates a SOSA result in RDF.}
\label{fig:py-create-sosa-result}
\end{subfigure}
\end{minipage}
\hfill
\begin{minipage}[t]{0.4\textwidth}
\begin{subfigure}[t]{\linewidth}
\begin{minipage}{\linewidth}
\begin{lstlisting}[language=Python,basicstyle=\ttfamily\scriptsize]
def get_qudt_unit_degree_celsius() -> URIRef:
    return QUDT_UNIT["DEG_C"]
\end{lstlisting}
\end{minipage}
\caption{Get the IRI for the QUDT unit.}
\label{fig:py-get-qudt-unit}
\end{subfigure}

\vspace{0.1em}

\begin{subfigure}[t]{\linewidth}
\begin{minipage}{\linewidth}
\begin{lstlisting}[language=Python,basicstyle=\ttfamily\scriptsize]
def create_oim_sosa_result_iri() -> URIRef:
    id = generate_unique_id()
    return OIM_RESULT[id]
\end{lstlisting}
\end{minipage}
\caption{Create an OIM result IRI.}
\label{fig:py-create-result-uri}
\end{subfigure}
\end{minipage}

\caption{Examples of low-level functions used in the sea temperature example.}
\label{fig:python-low-level-functions}
\end{figure*}

The ontology QUDT defines thousands of units of measure and quantity kinds. Each of them requires a low-level function that returns the appropriate IRI, such as the function \texttt{get\_qudt\_unit\_degree\_celsius()}. Developing and maintaining these functions by hand entails substantial effort and cost. In the next section, we introduce a data-driven approach for generating these functions, which significantly reduce the implementation cost.

\subsection{Data-Driven Generation of Python Functions Using Jinja Templates}
\begin{figure}[h!]
\centering
\includegraphics[scale=0.12]{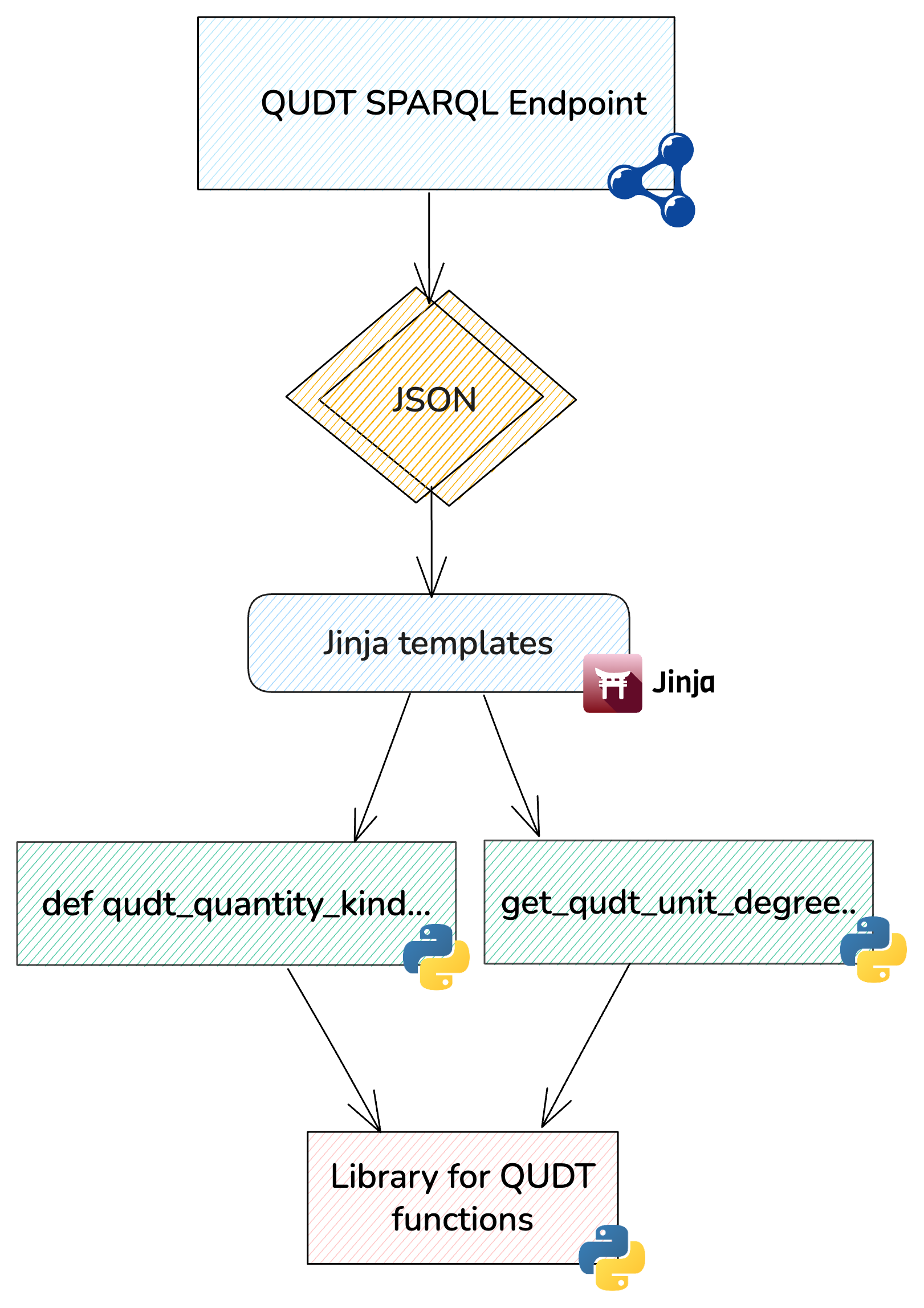}
\caption{Data pipeline for generating QUDT Python functions using Jinja templates.}
\label{fig:Jinja_generation_python_functions}
\end{figure}

As discussed earlier, the sheer number of quantity kinds and units of measure defined by the QUDT ontology, will require an equally large number of low-level Python functions. Manually implementing these Python functions would be labor-intensive, prone to inconsistencies, and difficult to sustain as the underlying vocabularies evolve. To address this, our framework incorporates a lightweight code-generation pipeline, ilustrated by Figure \ref{fig:Jinja_generation_python_functions}, that automatically produces Python functions directly from SPARQL query results, using Jinja as a flexible templating mechanism. This approach makes it possible to automatically generate thousands of functions for QUDT units, quantity kinds, SOSA observable properties, and OIM measurement types. It also allows effortless updates whenever external vocabularies change, ensuring long-term consistency. In addition, it maintains stable naming conventions across all generated functions. Finally, it removes repetitive boilerplate code so developers can focus on higher-level modelling logic.

The code generation pipeline takes the results of SPARQL queries and instantiates simple
Jinja templates to generate low-level functions. For example, a minimal Jinja template for low-level functions for units can be specified as follows:

\vspace{2mm}

\begin{lstlisting}[language=Jinja,basicstyle=\ttfamily\small]
{% for func in functions %}
def {{ func.name }}() -> {{ func.return_type }}:
    {% if func.docstring %}"""{{ func.docstring }}"""{% endif %}
    return {{ func.namespace }}["{{ func.constant }}"]
{% endfor %}
\end{lstlisting}

\vspace{2mm}

The first step in the pipeline uses SPARQL to query the QUDT ontology and retrieve all units together with their URIs and human-readable labels. A lightweight Python script then processes these SPARQL results to derive the fields expected by the Jinja template, such as the function name, the return type, the namespace prefix, and the local name used to construct the IRI. In this way, the SPARQL query itself remains simple, while the Python post-processing step bridges the gap between the raw query results and the structured input required by the template. The code generation pipeline then produces low-level Python functions of the form:

\vspace{2mm}
\begin{lstlisting}[language=Python,basicstyle=\ttfamily\small]
def get_qudt_unit_degree_celsius() -> URIRef:
    """Returns the IRI for QUDT unit: degree Celsius"""
    return QUDT_UNIT["DEG_C"]
\end{lstlisting}

\vspace{2mm}
Although it would be possible to expose units and other vocabulary terms through direct dictionary lookups on namespace objects, we intentionally generate named Python functions instead. Direct lookups would expose low-level ontology details (such as QUDT-specific identifiers) at higher abstraction levels, which conflicts with our goal of hiding such details from non-expert users. Named functions also provide a more discoverable and robust interface as autocompletion helps users find available units easily, and misspelled function names are typically caught immediately by Python.

Ontology engineers therefore only need to define the Jinja template specifications, while the code generation pipelines produce a large catalogue of low-level Python functions. This approach greatly improves maintainability and scalability, ensuring that the library remains synchronised with evolving semantic standards at minimal human cost.

\section{Related Work}
\label{related_work}
Semantic data harmonisation has been addressed through various approaches, each balancing expressiveness, automation, and usability differently. In Section \ref{use_case}, we presented two of these approaches, used directly in ILIAD, namely RML and OTTR, and we discussed why they were not finally adopted in the project. In this section, we review other technologies relevant to semantic data integration, though not adopted in our pilots. These include ontology design-pattern frameworks, template and mapping languages, programmatic APIs, and automation-oriented systems. Although each offers valuable contributions for modelling ontologies, none satisfy the requirements posed by ILIAD data scientists.

As discussed in Section \ref{use_case}, OTTR offers a conceptually elegant approach and a clear modelling discipline often absent in ad-hoc RDF generation. We therefore evaluated its suitability for ILIAD, but found that non-experts still face a steep learning curve due to required knowledge of OIM modelling details, OTTR syntaxes, and specialised tools. Python libraries such as pyOTTR \cite{pyottr} and maplib \cite{bakken2023maplib} reduce tooling overhead and integrate well into scientific workflows, yet still require substantial user effort.  Beyond OTTR and its associated Python libraries, several other template-based frameworks have been proposed. One example is Generic DOL (GDOL) \cite{krieg2017generic}, an extension of the Distributed Ontology, Modelling, and Specification Language (DOL) \cite{mossakowski2015distributed} that supports parameterised ontology templates across a wide range of logical formalisms. While more expressive than OTTR, it is also more complex, which makes it unsuitable for ILIAD data scientists.

Pattern-based ontology engineering is grounded in ontology design patterns (ODPs) \cite{hitzler2016ontology}, which offer reusable solutions to recurrent modelling challenges. ODPs are documented in repositories such as \texttt{OntologyDesignPatterns.org} \cite{gangemi2009ontology} and MODL \cite{shimizu2019modl}. Tools including XDP \cite{hammar2015ontology} and CoModIDE \cite{shimizu2019comodide} support pattern instantiation within WebProtégé. Although these methods promote high-quality ontology design, they rely on manual use and remain impractical for large-scale or automated generation of pattern instances, as required in the ILIAD project.

In Section \ref{use_case}, we reported our experience using RML, an RDF-based declarative mapping language for transforming relational, CSV, or JSON sources into RDF. Because RML’s syntax can be challenging for practitioners, researchers at Ghent University developed YARRRML \cite{heyvaert2018declarative} as a more user-friendly alternative. YARRRML provides a simpler YAML-based syntax for specifying RML rules, yet it still inherits the same challenges identified by ILIAD data scientists, namely, the need to understand detailed aspects of OIM and to learn additional tooling for RDF generation. Similar limitations also apply to other widely used mapping languages built on (or extending) the SPARQL query language, such as XSPARQL \cite{bischof2012mapping}, TARQL \cite{tarql}, and SPARQL-Generate \cite{lefrancois2017sparql}, as well as approaches based on Shape Expressions (ShEx) \cite{prud2014shape} like ShExML \cite{garcia2020shexml}, or JSON-driven specifications such as Helio \cite{cimmino2024helio}. Overall, although these solutions provide clear semantics and are suitable for data harmonisation pipelines, their syntactic complexity, limited reuse mechanisms, and distance from mainstream programming environments, such as Python, reduce their suitability for ILIAD’s scientific workflows.

Programming libraries in Java or Python, such as Apache Jena \cite{carroll2004jena}, the OWL API \cite{horridge2011owl}, RDFLib \cite{Krech_RDFLib_2025} and OWLready2 \cite{owlready2} provide fine-grained programmatic access to RDF graphs and OWL reasoning. These tools are mature and widely used, yet operate at a relatively low level: users must manually construct triples and manage modelling conventions. They do not provide built-in abstractions for reusable semantic patterns, making them unfit for ILIAD data scientist. 

There also exists a family of automation-oriented systems for knowledge graph construction. Materialisation engines such as Morph-RDB \cite{morph_rdb} and SDM-RDFizer \cite{iglesias2020sdm} execute R2RML/RML mappings, while virtualisation systems like Ontop \cite{ontop} expose relational data through SPARQL using R2RML-style specifications. These solutions are powerful, but they still require users to learn specialised mapping or query languages and do not integrate naturally into Python-based scientific workflows, making them difficult to adopt in ILIAD.

A separate line of work treats ontology engineering as software development. Tools such as Tawny-OWL \cite{lord2013semantic}, ROBOT \cite{jackson2019robot}, and KGTK \cite{ilievski2020kgtk} support modular, scriptable, and reproducible ontology or graph pipelines. While powerful, they rely on specialised languages and command-line tooling, and they do not map heterogeneous JSON or CSV data directly into ontology-aligned RDF within Python workflows. Consequently, they do not meet the practical needs of ILIAD data scientists who require tight integration with Python notebooks, DataFrames, and workflow orchestrators.

Overall, these approaches demonstrate the breadth of work on semantic modelling and transformation. Yet across the landscape, gaps remain for scenarios in which harmonisation must occur within Python, over heterogeneous JSON or API-based inputs, and in close interaction with DataFrames, notebooks, and workflow orchestrators. The functional, template-based design introduced in this paper specifically targets this space by providing executable abstractions grounded in
ontology design patterns while remaining native to the Python ecosystem.
\section{Conclusions and Future Work}
\label{conclusions_and_future_work}

This paper reported on practical experiences for semantic data harmonisation in the ILIAD project. Together with data scientists working in the project, we studied the suitability of several well-established technologies, including RML and OTTR. Although each tool provides significant advantages, our informal assessment showed that they align only partially with the working practices of ILIAD data scientists who operate primarily in Python environments, supported by Pandas DataFrames, notebooks, and data orchestration systems, such Dagster \cite{dagster}. The requirement to learn dedicated syntaxes, and specific tools, which exist outside their core programming environment introduced practical barriers that prevented their adoption.

In response, we developed a Pythonic functional approach to semantic data harmonisation that encodes SOSA, QUDT, and OIM design patterns directly as composable Python functions. These functions are organised into low-, mid-, and high-level abstractions to support clear separation of concerns and scalable reuse. At the lowest layer, functions capture OWL constructors and call methods provided by the Python library RDFLib. Mid-level functions specify ontology design patterns defined by the ontologies SOSA, QUDT and OIM, and they are implemented by calling low-level functions. Finally, high-level functions orchestrate domain-specific data harmonisation tasks by invoking mid-level functions. A Jinja-based code-generation mechanism further extends this approach by automatically producing thousands of unit and quantity-kind functions from SPARQL queries over QUDT, keeping the library synchronised with evolving vocabularies while avoiding manual boilerplate. To give an indication of scale: in the ILIAD Aquaculture pilot, the library comprised roughly 4{,}000 functions, the vast majority of which were low-level functions automatically generated from QUDT vocabularies using Jinja templates, with approximately 50 mid- and high-level functions implemented manually. Representative harmonisation runs produced approximately 6 million RDF triples, illustrating use at a non-trivial scale. 

Despite of the positive feedback from ILIAD data scientists, our approach comes with important limitations. First, it is intentionally Python-centric: the templates are executable code rather than language-independent specifications, which makes reuse from non-Python environments more difficult and reduces portability compared to standards such as RML or OTTR. Second, as the number of templates grows, careful versioning, documentation, and testing are required to avoid inconsistencies and to keep the function hierarchy understandable for new users.
This is also a problem for most template-based approaches, such as OTTR. Third, our functions representing ontology templates currently lack explicit typing and formal validation. This problem could be pragmatically addressed through tests, SHACL \cite{shacl2017} validation, and code review rather than through a dedicated formalism.

Future work will focus on consolidating and extending the framework along several directions. We plan to test our Pythonic approach beyond the project ILIAD to support other domains, such as energy and maritime. We would also like to evaluate the approach more systematically with data scientists and domain experts, measuring development effort, maintainability, and error rates compared with RML and OTTR as baselines. On the technical side, we intend to deepen the integration with validation mechanisms such as SHACL, for example by generating SHACL shapes alongside Python ontology templates, and to explore extensions for editors that provide ontology-aware type checks and autocompletion.

Finally, we see substantial opportunities for leveraging generative AI to support what we might call "agentic AI ontology engineering": an approach where developers provide high-level instructions in natural language, and AI agents help flesh out the corresponding templates, functions, and validation assets. Foundational Language Models can already draft harmonisation functions from ontology fragments, propose Jinja templates for repetitive low-level functions, and suggest accompanying unit tests or SHACL shapes from example data. Looking ahead, we plan to explore semi-automated generation of entire template layers directly from OIM modules and related ontologies, with human review guiding and validating the results. This direction could significantly reduce the effort required to maintain extensive Python template libraries while keeping them aligned with evolving semantic models, reinforcing our approach as a practical bridge between ontology engineering and data-science workflows.

\section*{Acknowledgements}
The authors express their sincere gratitude to the ILIAD data scientists and ontology engineers for their invaluable support throughout this work. The quality and clarity of this paper have further benefited from the insightful comments provided by the anonymous reviewers. This work has been funded by the European Commission projects ILIAD (No 101037643), CircularTwAIn (No 101058585) and TwinShip (No 101192583).

\bibliographystyle{plain}
\bibliography{references}

\end{document}